%Paper: 9203054
%From: LARUS@slacvm.slac.stanford.edu
%Date: Thu, 19 Mar 1992 17:55 -0800 (PST)

\input phyzzx

\def\gl{\lambda}
\def\p{\partial}
\def\phip{\p_+\phi}
\def\phim{\p_-\phi}
\def\phipm{\p_+\p_-\phi}
\def\rhop{\p_+\rho}
\def\rhom{\p_-\rho}
\def\rhopm{\p_+\p_-\rho}

\rightline {SU-ITP-92-12}
\rightline {March 1992}
\bigskip\bigskip
\title{Hawking Radiation and Back-Reaction\foot{Work supported in
part by NSF grant PHY89-17438.}}

\vfill
\author{Leonard Susskind and L\'arus Thorlacius}
\bigskip
\address{ Department of Physics \break Stanford University,
            Stanford, CA 94305}
\vfill

\abstract
\singlespace
The puzzles of black hole evaporation can be studied in the
simplified context of 1+1 dimensional gravity.
The semi-classical equations of Callan, Giddings, Harvey and
Strominger provide a consistent description of the
evaporation process which we describe in detail.
We consider the possibility that black hole evolution leads
to massive stable remnants.  We show that such zero temperature
remnant solutions exist but we also prove that a decaying black
hole cannot evolve into one of them.
Finally we consider the issue of loss of quantum information behind
the global event horizon which develops in these geometries.  An
analogy with a well known solvable system shows that there may be
less to information than meets the eye.

\vfill\endpage

\REF\hawi{S.~W.~ Hawking
\journal Comm .Math. Phys. & 43 (75) 199.}
\REF\hawii{S.~W.~Hawking
\journal Phys. Rev. & D14 (76) 2460.}
\REF\thooft{G. 't Hooft
\journal Nucl. Phys. & B335 (90) 138, and references therein.}
\REF\cghs {C.~G.~Callan, S.~B.~Giddings, J.~A.~Harvey and
A.~Strominger, {\it Evanescent black holes}, preprint UCSB-TH-91-54,
EFI-91-67, PUPT-1294, November 1991.}
\REF\chrful{S.~M.~Christensen and S.~A.~Fulling
\journal Phys. Rev. & D15 (77) 2088.}
\REF\bddo{T.~Banks, A.~Dabholkar, M.~R.~Douglas and M.~O'Loughlin,
{\it Are Horned Particles the Climax of Hawking Evaporation?},
Rutgers University preprint, RU-91-54, January 1992.}
\REF\rst{J.~G.~Russo, L.~Susskind and L.~Thorlacius, {\it Black
Hole Evaporation in 1+1 Dimensions}, Stanford University preprint,
SU-ITP-92-4, January 1992.}
\REF\ac{Y.~Aharonov, A.~Casher and S.~Nussinov
\journal Phys. Lett. & 191B (87) 51.}
\REF\gibma{G.~W.~Gibbons and K.~Maeda
\journal Nucl. Phys. & B298 (88) 741.}
\REF\ghs{D.~Garfinkle, G.~Horowitz and A.~Strominger
\journal Phys Rev & D43 (91) 3140.}
%\REF\wald{R.~Wald, {\it General Relativity}, The University of
%Chicago Press, 1984.}
\REF\pol{A.~M.~Polyakov
\journal Phys. Lett. & 163B (81) 207.}
\REF\steve{S.~B.~Giddings, {\it Black Holes and Dark Remnants},
UCSB preprint, in preparation.}
\REF\bghs{B.~Birnir, S.~B.~Giddings, J.~A.~Harvey and A.~Strominger,
{\it Quantum Black Holes}, preprint UCSB-TH-92-08, EFI-92-16,
March 1992.}
\REF\hawiii{S.~W.~Hawking, {\it Evaporation of Two Dimensional Black
Holes}, Caltech preprint, CALT-68-\#1, March 1992.}
\REF\unruh{W.~G.~Unruh
\journal Phys. Rev & D14 (76) 870.}
\REF\unwald{W.~G.~Unruh and R.~M.~Wald
\journal Phys. Rev. & D29 (82) 1047.}
\REF\witten{E.~Witten
\journal Phys. Rev. & D44 (91) 314.}
\REF\alstr{M.~Alford and A.~Strominger, {\it S-Wave Scattering of
Charged Fermions by a Black Hole}, preprint NSF-ITP-92-13, February
1992.}

\chapter{Introduction}

Black hole evaporation provides us with an interesting set of puzzles
where quantum mechanics and general relativity clash
[\hawi,\hawii,\thooft].  The central paradox is that matter in a
quantum mechanically pure state can undergo gravitational collapse
to form a black hole, which then evaporates, and in this process the
initial pure state appears to evolve into a mixed quantum state,
describing the outgoing Hawking radiation.  This argument is based on
a calculation which treats matter quantum mechanically in a background
geometry which evolves according to Einstein's classical field
equations.  It seems prudent to include the perhaps subtle effects of
back-reaction on the geometry, due to the emission of Hawking
radiation, before we start tinkering with quantum mechanics.  Hopefully
these issues can be resolved without direct reference to Planck scale
physics (after all the horizon of a massive black hole forms in a
region of weak gravitational coupling) allowing us to explore
interesting features of quantum gravity without having to grapple
with a theory of everything.

It is often illuminating to address difficult physical problems in
the simplified context of 1+1 dimensions.  A particularly interesting
toy model for black hole physics was suggested by Callan, Giddings,
Harvey and Strominger (CGHS) in reference [\cghs], where they coupled
two-dimensional gravity to a dilaton and conformal matter fields.
CGHS described a set of exact classical solutions, including black
holes formed by infalling shock waves, and they observed that
Hawking radiation can be described by adding correction terms to the
classical equations of motion.  These extra terms derive from the
conformal anomaly of the matter fields [\chrful] and are local in
conformal gauge, with the usual non-locality appearing only in the form
of boundary conditions required to satisfy constraints.  CGHS also
pointed out that once these quantum corrections have been included
one can study the back-reaction on the geometry, due to the Hawking
radiation, using classical methods in the effective theory.  They
argued that the quantum corrections would prevent the usual black
hole singularity from forming in gravitational collapse and therefore
quantum information would not be lost.  Subsequently it was shown
that in this theory gravitational collapse always develops a
singularity [\bddo,\rst].  Thus the paradox has not been resolved.

In this paper we study the CGHS model further.
Background material can be found in references [\cghs], [\bddo]
and [\rst].
We begin in section~2
by briefly reviewing the model and describing some exact classical
solutions.  We then turn to the quantum theory in section~3.  Following
CGHS we study Hawking radiation and its back reaction on black holes
using effective equations of motion, with correction terms induced
by the conformal anomaly of the matter fields.
Using both analytic
and numerical methods we are able to present a fairly
detailed picture of the formation and quantum mechanical evaporation
of 1+1 dimensional black holes.

An important issue to settle is
the nature of the final state that a 1+1 dimensional black hole will
settle down to.  One can think of a number of possibilities, all of
which are problematic to some degree.  The first, suggested by
Hawking [\hawii], is that the black hole completely evaporates
leaving no remnant behind but only incoherent radiation, and quantum
coherence is lost in the process.  This is a radical proposal and
more conservative alternatives should be exhausted before we modify
quantum mechanics.

Another possibility is that the Hawking process leaves behind a
stable remnant whose precise quantum state reflects the initial
configuration of the system [\ac].  Evidently this requires a
virtually infinite number of distinct quantum states of black hole
cadavers.  There are two very different cases to consider.  On the
one hand the mass and size of the remnants would be of order the
Planck scale, in which case their density of states must be infinite.
If these light remnant states couple to gravity in any conventional
way, they would be copiously produced in pairs whenever there is
available a Planck scale amount of energy.  The energy would not
have to be very concentrated because even if the probability for
producing a particular pair of these Planck scale objects is tiny
the infinite density of states makes the total rate diverge.  For
example, the sun would have been short lived indeed if such channels
had been available.
It has recently been suggested that the black hole remnants
cannot be produced in pairs [\bddo] but this in itself is a
radical departure from the principles of relativistic quantum mechanics.

Alternatively the Hawking process might terminate while the black hole
still has macroscopic mass.  That could only happen if the
back-reaction on the geometry develops a singularity on the apparent
horizon itself.  This may sound outlandish but in section~3 we will
show that the curvature on the apparent horizon does in fact increase
as the black hole evaporates and in section~4 we exhibit static,
non-radiating solutions of the CGHS-equations which are candidates
for such remnants.  Massive remnants can carry a lot of information
and their density of states would not have to diverge in order to
preserve quantum coherence, so this possibility seems worth exploring.
However, we are able to prove that it is not realized in 1+1 dimensional
gravitational collapse, as described by the CGHS-equations.

Perhaps the most conservative option is that despite its thermal
appearance Hawking radiation really carries off information about the
quantum state of the black hole [\thooft].  This would reconcile the
evolution of black holes with conventional quantum mechanics.  In
section~5 of this paper we will discuss to what extent this type of
questions can be addressed in this 1+1 dimensional model, but
unfortunately we cannot offer any definitive conclusions at present.

\chapter{The Model}

The system we will consider is two-dimensional gravity coupled to a
dilaton $\phi$ and a set of scalar fields $f_i$.  The classical
dynamics is governed by the action
$$
S = {1\over 2\pi} \int d^2x\>\sqrt {-g} \bigl[
  e^{-2\phi}(R+4(\nabla           \phi )^2
+4\gl ^2)-{1\over 2}\sum_{i=1}^N (\nabla f_i)^2 \bigr]\ ,
\eqn\lagr
$$
which resembles the target space action of $c{=}1$ non-critical
string theory.  It can be viewed as an effective action for radial
modes of four-dimensional extreme dilaton black holes
[\cghs,\gibma,\ghs] but in this paper we will study it on its own
merits as a two-dimensional model of gravity coupled to matter.

The matter fields $f_i$ are minimally coupled to the metric and do
not directly interact with the dilaton.  We will assume there is a
large but finite number of them.  The reason for this is that the
one-loop quantum corrections to the action are dominated by the
conformal anomaly of the matter fields if $N$ is large and the
effective action takes a particularly simple form.  The arbitrary
parameter $\gl$ defines a mass scale in this theory and we will
refer to a black holes with $M>N\gl$ as massive.

The classical equations of motion obtained from the action \lagr\ are
$$\eqalign{
\nabla ^2 f_i =&\> 0 \, ,  \cr
R + 4\gl^2 + 4\nabla ^2\phi - 4(\nabla \phi)^2 =& \> 0 \, ,  \cr
2e^{-2\phi} \bigl[ \nabla _\mu \nabla _\nu \phi
 + g_{\mu\nu}\bigl( (\nabla \phi)^2-\nabla ^2\phi -\gl^2 \bigr)\bigr]
 =& {1\over 2} \sum_{i=1}^N \bigl( \nabla _\mu f_i \nabla _\nu f_i
  -{1\over 2} g_{\mu\nu} (\nabla f_i)^2 \bigr) \, . \cr}
\eqn\coveom
$$
It is convenient to work in conformal gauge
$$\eqalign{
g_{+-} = g_{-+} =& -{1\over 2} e^{2\rho} \, ,  \cr
g_{--} = g_{++} =& \> 0 \, ,                      \cr}
\eqn\cgauge
$$
with light-cone coordinates $x^{\pm} = x^0 \pm x^1$.  The fields
in the theory are then $f_i,\phi$ and $\rho$, and the classical
equations of motion become,
$$\eqalign{
\p_+\p_- f_i =&\, 0\,,   \cr
2 \phipm - 2 \phip \phim -{\gl^2\over 2} e^{2\rho} =&\,\rhopm\,,\cr
\phipm - 2\phip\phim - {\gl^2\over 2} e^{2\rho} =&\, 0 \,, \cr}
\eqn\cgeom
$$
respectively.  In addition, one must impose as constraints the
equations of motion corresponding to the metric components, which
are set to zero in this gauge:
$$\eqalign{
e^{-2\phi}(2\p_+^2\phi -4 \rhop\phip ) =&
   {1\over 2} \sum_{i=1}^N \p_+f_i \p_+f_i \, , \cr
e^{-2\phi}(2\p_-^2\phi -4 \rhom\phim ) =&
   {1\over 2} \sum_{i=1}^N \p_-f_i \p_-f_i \, . \cr}
\eqn\constr
$$

A number of exact solutions of these equations can easily be found
[\cghs].  The simplest one is the vacuum solution
$$\eqalign{
f_i =&\> 0 \,,  \cr
e^{-2\phi} = e^{-2\rho} =& -\gl^2 \,x^+ x^- \, . \cr}
\eqn\vac
$$
This is just the `linear dilaton' background of non-critical string
theory.  A change of variables, $x^\pm = \pm e^{\pm u^\pm}$,
makes the metric flat and the dilaton field linear in $u^+{-}u^-$.
A `static' black hole is described by
$$\eqalign{
f_i =& 0 \,,  \cr
e^{-2\phi} = e^{-2\rho} =& {M\over \gl} -\gl^2 \,x^+ x^-\,. \cr}
\eqn\static
$$
Finally one can describe the metric and dilaton fields
due to an infalling shell of matter by patching together
the vacuum solution and a black hole solution across some light-like
line $x^+ = x_0^+$ (see figure 1),
$$
e^{-2\phi}=e^{-2\rho}=
\cases{\phantom{33333} -\gl^2 x^+x^- &if $x^+<x^+_0$; \cr
-{M\over \gl x^+_0}(x^+-x^+_0) -\gl^2 x^+x^- &if $x^+>x^+_0$. \cr}
\eqn\infal
$$
Note that $\phi$ and $\rho$ are continuous across the matching line.
Inserting $\phi$ and $\rho$ from \infal\ into the ++ constraint
equation in \constr\ reveals that this solution is indeed the
response to an incoming shock wave with energy-momentum
${1\over 2}\sum_{i=1}^N \p_+f_i\p_+f_i
={M\over \gl x_0^+}\delta (x^+-x^+_0)$.
The geometry is singular on the curve where $e^{-2\rho}=0$ and
the line $x^- = - {M\over \lambda^3 x^+_0}$ is a global event
horizon in this classical solution.

The factor of $e^{-2\phi}$ in front of the gravity
terms in the action \lagr\ makes the strength of gravitational
quantum corrections depend on the magnitude of the dilaton field,
$g \sim e^{\phi}$.
Both the linear dilaton vacuum and the classical black hole
solutions have a position dependent coupling strength with weak
coupling being asymptotically far away.  In fact, the strength of
quantum corrections provides a coordinate invariant notion of
position in this one-dimensional world.  Recall that in the
four-dimensional Schwarzschild solution lines of constant radial
distance (and fixed angle) go from being time-like outside the
horizon to being space-like inside the black hole.  The analogous
statement for our 1+1 dimensional black holes is that
$(\nabla\phi)^2$ changes sign at the horizon [\rst].  This
provides us with a sensible local definition of an apparent
horizon which will be useful in what follows.

\chapter{Quantum Corrections}

The most important result of the original paper of CGHS [\cghs] is
that the back-reaction on the geometry due to Hawking radiation can
be described by adding appropriate correction terms to the classical
equations of motion.
When there is a large number of matter fields in the theory there
is a class of quantum corrections, associated with closed matter
loops, which become important when $g^2N\sim 1$, long before other
gravitational corrections kick in.  At the one-loop level the matter
conformal anomaly is the only term of this class and it will be
responsible for the quantum back-reaction we will explore later on.
When a black hole is formed the coupling at
the horizon is $g^2 \sim {\gl\over M}$ so our approximations remain
valid so long as the mass remains large compared to $N\gl$.

At one-loop order the anomaly contributes the well known non-local term
$$
-{N\over 96\pi} \int d^2x\>\sqrt{-g}\,R\,{1\over \nabla^2}\,R
\eqn\nonloc
$$
to the effective action [\pol].\foot{Following CGHS we assume that
the Liouville cosmological constant term can be fine tuned to zero.}
In the conformal gauge \cgauge\ this term appears local,
$-{N\over 12\pi} \int d^2x \> \rhop\rhom$ and
the one-loop corrected equations of motion, which we refer to as
the CGHS-equations, can be arranged as follows,
$$\eqalign{
\p_+\p_-f_i =&\,0,      \cr
\phipm =&\, (1{-}{N\over 24}e^{2\phi}) \, \rhopm \, ,   \cr
  2\, (1{-}{N\over 12}e^{2\phi}) \,\phipm =&\,
  (1{-}{N\over 24}e^{2\phi}) \,
  (4\phip \phim + \lambda^2 e^{2\rho}) \, .  \cr}
\eqn\eom
$$
The constraint equations also receive corrections,
$$\eqalign{
e^{-2\phi}(2\p_+^2\phi - 4\rhop\phip ) =&\,
{1\over 2}\sum_{i=1}^N \p_+f_i\p_+f_i - {N\over 12}
\bigl(\rhop\rhop - \p_+^2\rho + t_+(x^+) \bigr) \, , \cr
e^{-2\phi}(2\p_-^2\phi - 4\rhom\phim ) =&\,
{1\over 2}\sum_{i=1}^N \p_-f_i\p_-f_i - {N\over 12}
\bigl(\rhom\rhom - \p_-^2\rho + t_-(x^-) \bigr) \, . \cr}
\eqn\nconstr
$$
The functions $t_\pm(x^\pm)$ are needed to satisfy
asymptotic physical boundary conditions [\cghs], so
there remains a degree of non-locality in the conformal gauge.
As expected the quantum corrections in these equations appear with
factors of $N e^{2\phi}$ compared to the classical terms.

For some purposes it can be useful to have a description of this
system outside conformal gauge.  In a more general gauge the
anomaly induced term \nonloc\ can be made local by adding a new
degree of freedom which is coupled to the scalar curvature,
$$
S_Z = {1\over 2\pi} \int d^2x\>\sqrt {-g} \bigl[
        -{1\over 2}(\nabla Z)^2 + \sqrt{N\over 24}Z R\bigr]\ .
\eqn\zaction
$$
The non-local form \nonloc\ is recovered when $Z$ is integrated out
of the path integral.  The energy momentum tensor of the $Z$ field is
$$
T_{\mu\nu} = {1\over 2}\bigl[\nabla_\mu Z \nabla_\nu Z - {1\over 2}
              g_{\mu\nu} (\nabla  Z)^2\bigr]
            + \sqrt{N\over 24} \bigl[\nabla_\mu \nabla_\nu Z -
              g_{\mu\nu} \nabla^2Z\bigr] \,.
\eqn\zemt
$$
The conformal anomaly of the matter fields
appears as a non-vanishing trace of the classical energy-momentum
tensor of $Z$.
This is analogous to what happens in bosonization of two-dimensional
spinor electrodynamics where the axial anomaly of the fermions can be
expressed classically in terms of the bosonizing field.  We will
develop this analogy further in section~5 when we discuss
the fate of quantum information in this theory.

The classical equations \cgeom\ can be made linear by a combination
of gauge choices and field redefinitions, which makes them
relatively easy to solve.  Unfortunately these tricks no longer
work when the one-loop corrections have been added, and we now have
coupled, non-linear, partial differential equations to
deal with.  It is easily checked that the linear dilaton vacuum
\vac\ remains an exact solution.  The geometry due to an incoming
shock wave can no longer be found in closed form but is described by
some non-trivial solution of the CGHS-equations which is matched
continuously onto the vacuum across $x^+=x^+_0$.  It approaches a
classical black hole solution asymptotically far away, where the
coupling is weak.  In the remainder of this section we will use
the equations of motion and constraints to explore qualitative
features of such solutions, reinforcing and extending the
observations of reference [\rst].  We will also present numerical
calculations of two-dimensional black hole evolution which support
our analytic results.

The first thing to note is that a curvature singularity forms on
the infall trajectory at $e^{2\phi}={12\over N}$.  This was shown
in [\bddo] and [\rst] and we will not repeat the argument here.
The singularity extends into a curve in the
$x^+>x^+_0$ region through which physical solutions cannot be
meaningfully extended.  Our approximations break down as the
singularity is approached and it may or may not persist in the full
quantum theory, but its presence at this level means that care must
be taken in developing any systematic approximations.  Perhaps one
should view the ${N\over 12}e^{2\phi}=1$ line as a natural boundary
to spacetime in the quantum theory.  We will at any rate focus
on the evolution of massive black holes where the CGHS-equations are
reliable so long as the mass remains large compared to $N\gl$.
Our results are for the most part independent of how the singularity
issue is ultimately resolved.  It is worth pointing out that removing
the singularity by modifying the short distance physics will not
automatically cure problems of information loss.  The evaporation
process would still be governed by the semi-classical equations for
most of the black hole lifetime and there will be a limited amount
of energy left to carry off the information contained in the original
black hole (we are assuming for the moment that it does not get
radiated away during the semi-classical evaporation).  Either it
would be contained in a light stable remnant or have to be emitted
in very low energy radiation over a long time.  In either case we
would have to contend with a virtually infinite density of
(meta-)stable states at the Planck scale in the quantum theory.

The singularity is cloaked by an apparent horizon which forms at
$x^- \simeq -{M\over \gl^3 x^+_0}$, for $M$ large, and then begins
to recede as the black hole emits energy in Hawking radiation [\rst].
Later on we will show that the apparent horizon continues to recede
as long as there is outgoing energy flux from the black hole.  This
intuitively obvious result provides a nice consistency check on the
picture of black hole evaporation presented by the CGHS-equations.

The picture of black hole evaporation which emerges is qualitatively
as follows.  The curve of singularity at ${N\over 12}e^{2\phi}=1$
forms inside the apparent horizon where $\nabla \phi$ is time-like
and lines of constant $\phi$ are space-like.  Thus it follows that
the line of singularity is everywhere space-like unless it were to
cross the apparent horizon.  Such a crossing, however, can not happen
since $\phip$ equals zero on the apparent horizon but diverges on the
singularity.  Once the black hole has formed it emits Hawking
radiation and the apparent horizon recedes.  Since it cannot cross
the singularity they must both approach light-like lines.  We will
present analytic and numerical evidence which strongly suggests that
they in fact approach the same light-like line from opposite sides as
in figure~2.  The light-like asymptote is a global event horizon.

If all the mass of the black hole is radiated then one expects the
geometry to approach the vacuum solution.  Alternatively the
black hole might approach a massive stable remnant whose properties
would emerge at asymptotically late times.
To investigate the asymptotic behavior of black hole
solutions one can consider
how coordinate invariants, such as the curvature scalar or
$(\nabla \phi)^2$, evolve along contours of constant coupling strength.
One can, for example, derive a simple relation
between the rate of change of $R$ along a curve of constant $\phi$
and the flux of energy across the curve.  For this purpose it is
convenient to introduce local coordinates $u,v$ in the neighborhood
of the contour, such that $u$ parametrizes the contour,
$\phi\bigl(\tilde x^+(u),\tilde x^-(u)\bigr)\equiv\phi_0$,
and $v$ parametrizes orthogonal curves.  A good choice is
to take $v$ as $\phi$ itself and normalize $u$ in such a way that
the Jacobian determinant of the coordinate transformation from
$x^+,x^-$ to $u,v$ equals one.  The matter energy, including Hawking
radiation, in this coordinate system is $T^{u u}$ and the outgoing
energy flux is $T_u^{\phantom{u}v}$.

The curvature scalar in the conformal gauge \cgauge\ is given by
$$
R = 8 e^{-2\rho} \rhopm \,,
\eqn\curv
$$
which can be reexpressed using the equations of motion \eom\ as
$$
R = {4\over 1-{N\over 12}e^{2\phi}}
\bigl(4e^{-2\rho}\phip\phim + \gl^2 \bigr)  \,.
\eqn\cur
$$
We are interested in the rate of change of curvature along the
$\phi=\phi_0$ contour,
$$\eqalign{
\dot R \equiv & {d\over du}\,R \big\vert_{\phi=\phi_0}  \cr
       =& {\p \tilde x^+\over \p u}\, \p_+ R
         +{\p \tilde x^-\over \p u}\, \p_- R            \cr
       =& \p_-\phi \,\p_+ R - \p_+\phi \,\p_-R   \>.    \cr}
\eqn\rdota
$$
Inserting the expression \cur\ for the curvature gives
$$
\dot R = {8e^{-2\rho}\over 1{-}{N\over 12}e^{2\phi_0}} \bigl[
         (\p_-\phi)^2(2\p_+^2\phi{-}4\rhop\phip)
       - (\p_+\phi)^2(2\p_-^2\phi{-}4\rhom\phim) \bigr] \,.
\eqn\rdotb
$$
The constraint equations \nconstr\ reveal that
$e^{-2\phi}(2\p_\pm^2\phi{-}4\p_\pm\rho\p_\pm\phi)$ is the
gravitational response to the energy-momentum tensor of matter and
Hawking radiation so we can write
$$
\dot R = {8e^{-2\rho+2\phi_0}\over 1{-}{N\over 12}e^{2\phi_0}} \bigl[
         (\p_-\phi)^2\,T_{++} - (\p_+\phi)^2\,T_{--} \bigr] \,.
\eqn\rdotc
$$
The final step is to observe that the expression in the square
brackets is proportional to the energy flux across the contour and
we obtain the desired result
$$
\dot R = -{4e^{2\phi_0}\over 1{-}{N\over 12}e^{2\phi_0}}\,
T_u^{\phantom{u}v} \>.
\eqn\rdot
$$
This relationship can be checked by considering a contour in the
classical region $Ne^{2\phi_0}<<1$.  The value of the curvature of the
classical solution \infal\ where the contour meets the infall line is
$R=4\gl e^{2\phi_0} M$.  If the curvature on the contour goes to
zero at late times then \rdot\ implies that the integrated flux
of Hawking radiation carries off an amount of energy equal to the
original mass.  In other words, the geometry approaches flat space if
and only if the black hole radiates away all its energy.

We are now ready to show that the apparent horizon always recedes
if there is any outgoing flux of energy from the black hole.  Let the
line of apparent horizon be parametrized as $x^-=\hat x^-(x^+)$.
By definition $\phip$ vanishes everywhere along this curve,
$$
0 = {d\over dx^+}\, \phip \Bigl\vert _{x^-=\hat x^-}
  = \p_+^2 \phi + ({d\hat x^-\over dx^+})\, \phipm \,.
\eqn\hline
$$
The constraints \nconstr\ relate the value of $\p_+^2\phi$ on
the apparent horizon to the outgoing energy flux at that point.
Using this and the second equation of motion in \eom\ we can write
the slope of the horizon curve in terms of the flux as follows,
$$
{d\hat x^-\over dx^+} =
 {e^{2\phi}(1{-}{N\over 12}e^{2\phi})
 \over 2\gl^2 (1{-}{N\over 24}e^{2\phi})(\phim)^2}\>
 T_u^{\phantom{u}v} \,.
\eqn\slope
$$
The horizon recedes at a rate proportional to the outward flow of
energy from the black hole (and advances when the black hole absorbs
infalling matter).

On the apparent horizon the expression for the curvature \cur\ reduces
to $R={4\gl^2 \over (1{-}{N\over 12}e^{2\phi})}$ and it immediately
follows that the horizon curvature increases as the black hole
evaporates and the apparent horizon recedes to larger values of
$e^{2\phi}$.

The apparent horizon of a massive black hole forms in a region of
weak coupling, $e^{2\phi}\simeq {\gl\over M}$,
and a good approximation to
${d\hat x^-\over dx^+}$ is obtained by inserting the classical
solution \infal\ on the right hand side in \slope .
To leading order this gives [\rst]
$$
{d\hat x^-\over dx^+} = {N\over 48\lambda^2{x^+_0}^2} \,,
\eqn\appslop
$$
which corresponds to a
rate of energy loss of ${N\lambda^2\over 48}$ in agreement with
the asymptotic flux of Hawking radiation found by CGHS [\cghs].
So long as the remaining black hole mass is large compared to
$N\gl$ the rate of energy loss is slow compared to the mass and
\appslop\ (with $x^+_0$ replaced by $x^+$) gives the instantaneous
slope.  Integration gives an 'adiabatic' approximation to the
horizon curve,
$$
\hat x^-(x^+) = \hat x^-(x^+_0) + {N\over 48\gl^2} \Bigl(
                  {1\over x^+_0} - {1\over x^+} \Bigr) \>.
\eqn\xmad
$$
In the $x^+\rightarrow\infty$ limit this approaches a global
horizon at $\hat x^- = \hat x^-_0 + {N\over 48\gl^2 x^+_0}$.
For a very large initial mass this should be an excellent estimate
because \appslop\ is valid for most of the black hole lifetime.
The approximation breaks down when the coupling strength becomes
too large.  Adiabatic arguments similar to those given above lead
to a linearly growing coupling strength at the apparent horizon,
$$
e^{2\phi} = {\gl x^+\over M x^+_0} \>,
\eqn\coupad
$$
which remains small for a long time if the original mass is large.

Note that in the $x^\pm$ coordinate system the total distance along
$x^-$, which the apparent horizon recedes, is independent of the
black hole mass.  This means that the global event horizon extends
far into the weak coupling region and at first sight it appears that
some, if not almost all, of the information about the quantum state
of the system will be lost behind it.  However, as we shall soon see,
the back-reaction is strong enough to convert all the energy of the
infalling matter to outgoing Hawking radiation by the time the
incoming shock wave reaches the global event horizon.  We will address
the question of information loss in section~5 and at that point it
will be useful to have a clear notion of the total energy carried
across light-like lines of constant $x^-$, such as the global horizon.

We will define the total energy on a given $x^-$ line as
$$
M(x^-) = \lim_{x^+\rightarrow\infty} {1\over 4\gl}
         (1-{N\over 12}e^{2\phi})^{3\over 2} \,e^{-2\phi}\, R \>.
\eqn\mtot
$$
This definition gives the correct mass of classical black hole
solutions.  The factor of $(1-{N\over 12}e^{2\phi})^{3\over 2}$
does not affect this since it goes to 1 as $x^+\rightarrow\infty$
but it is convenient for what follows.  For large negative $x^-$
this expression for $M(x^-)$ gives the standard initial total
energy of the infalling matter in gravitational collapse.  As
$x^-$ increases the remaining mass will be reduced by an amount
equal to the energy that has been radiated out in Hawking radiation up
to that point.  Using the equations of motion and the constraints one
obtains
$$
{d M\over dx^-} = {2\over \gl} \phip \,e^{-2\rho}\, T^m_{--} \, ,
\eqn\dmdx
$$
where $T^m_{--}$ is the -- component of the matter energy momentum
tensor with the contribution from the anomaly included.  The
factors in front of $T^m_{--}$ reflect the fact that the
$x^+,x^-$ coordinate system is not asymptotically Minkowskian.
Thus \mtot\ appears to be a sensible definition of the mass
and it will clearly go to zero at the global event horizon if all
the energy of the black hole is radiated away.  $M(x^-)$ can be
written as an integral as follows,
$$
M(x^-) = {1\over 4\gl}\int_0^\infty dx^+ \,{\p \over \p x^+}\bigl[
         (1-{N\over 12}e^{2\phi})^{3\over 2} e^{-2\phi} R \bigr] \,,
\eqn\massint
$$
and by using the equations of motion and the constraints this can be
worked into,
$$
M(x^-) = {1\over 4\gl}\int_0^\infty dx^+\,\sqrt{1-{N\over 12}e^{2\phi}}
              e^{-2\rho} \phim T^m_{++} \>.
\eqn\mint
$$
We will make use of this expression in section~5.

We close this section by presenting some preliminary numerical
results on black hole evolution in this 1+1 dimensional model.  We
consider a black hole formed by an incoming shock wave.  When
viewed on lines of constant $x^+$ the equations of motion \eom\ are
ordinary differential equations for $\phip$ and $\rhop$,
$$\eqalign{
\p _-(\phip) = & (1-{N\over 24}e^{2\phi}) \p _-(\rhop)\,,   \cr
2(1-{N\over 12}e^{2\phi}) \p _-(\phip) = & (1-{N\over 24}e^{2\phi})
            (4\phim \phip + \gl^2 e^{2\rho})\,.    \cr}
\eqn\ordeq
$$
This was used in reference [\rst] to obtain exact information about
the solution just above the infall line $x^+=x^+_0$ and to
establish the formation of the singularity at
${N\over 12}e^{2\phi}=1$.  The equations \ordeq\ can be solved
exactly on the infall line because the functional coefficients in
them are explicitly known there in terms of the linear dilaton
background.  But now the information about $\phip$ and $\rhop$ on
the infall line can be used to estimate the solution on the line
$x^+=x^+_0+\epsilon^+$ where $\epsilon^+$ is some small step size.
Then the ordinary differential equations \ordeq\ can be solved
numerically for $\phip$ and $\rhop$ on that line, making it
possible to take another step in the $x^+$ direction.  Figure~3
shows typical results of evolving a gravitational collapse solution
of the CGHS-equations in this way.  The numerical integration cannot
get through the singularity but since it lies on a space-like curve
this does not affect the numerical solution in the physical region.
The apparent horizon recedes as expected and it appears to approach
the singularity asymptotically.

We have also located some contours of constant $\phi$ (which are
not shown in figure~3) and evaluated the curvature scalar along them.
The curvature does decrease along these contours as expected from
\rdot .  Numerical instability prevents us from following the contours
very far into the $x^+$ direction and we can not at present determine
from our data whether $R$ goes to zero asymptotically.  The hyperbolic
nature of the CGHS-equations will make most numerical approaches
unstable but one can certainly do better with a more sophisticated
numerical method that the one we have employed here.

\chapter{Static Black Holes}

In this section we will discuss some properties of static black
holes and speculate about their possible role as massive stable
remnants of black hole evaporation.  Some related issues are
discussed in [\steve]
We will present numerical
evidence for the existence of static, non-radiating solutions of
the CGHS-equations, which are candidates for stable 1+1 dimensional
remnants.\foot{These numerical solutions have also been obtained by
J. Russo.  While preparing this paper for publication we
learned that B.~Birnir, S.~Giddings, J.~Harvey and A.~Strominger
have found numerical static solutions to the CGHS-equations
[\bghs] and that they have also been studied by
S.~Hawking [\hawiii].}
These solutions exist for any value of black
hole mass and they have some interesting properties.  In particular,
the coupling is weak, $Ng^2<<1$, everywhere in space if the mass is
large, and thus it is possible that all further quantum corrections
to these solutions remain small.

At large positive $x$ the static solutions rapidly tend to their
classical counterparts of the same mass but there is no energy in
Hawking radiation in this region.  For a massive black hole,
$M>>N\gl$, the classical geometry is well approximated until one
approaches the position of the classical horizon, but there, in a
metrically small region, the metric and dilaton change radically.
In particular the curvature first increases to a maximum value,
which depends on the mass, and then rapidly decreases, changes sign
and goes to minus infinity at a finite proper distance, as one
advances in the minus $x$ direction.
More surprisingly the dilaton reaches a maximum value at the same place
as the curvature and thereafter the coupling strength
decreases monotonically to zero as one approaches the curvature
singularity.  The maximum coupling is always smaller than the coupling
in the classical solution at the horizon, {\it i.e.} $g^2<{\gl\over M}$.
This means that one might not expect the solution to receive any
large corrections if $M>>N\gl$.

As mentioned in the introduction, massive remnants of this sort offer
the possibility of carrying off information about the quantum state
of a black hole without leading to an infinite density of states at
the Planck scale.  The naked curvature singularity might pose
problems of its own but these solutions are not ruled out on any
{\it a priori} grounds.  Later on we will consider the question of
whether black hole evaporation can lead to one of these massive remnants
as a final state.  The answer turns out to be negative, at least in
1+1 dimensions, for we are able to show that the gravitational collapse
solutions of the CGHS-equations cannot approach massive static black
holes asymptotically.  This of course does not preclude the possibility
of tunneling into one of these states at some stage in the evolution,
but that process is presumably suppressed by large barrier factors for
macroscopic black holes.

Before we can rule out the static solutions as the final states, we
need to explore some of their properties.  We begin with a brief
digression on the quantum theory of a free field in Rindler
space [\unruh,\unwald].  This simple system shares important features
with black holes both in four and two spacetime dimensions.

The $\vert x\vert > |t|$, $x>0$ wedge of flat two-dimensional Minkowski
space may be described by Rindler coordinates $R,\theta$ as follows,
$$\eqalign{
t =& R \sinh{\theta} \, , \cr
x =& R \cosh{\theta} \, , \cr}
\eqn\rindler
$$
or $x^\pm =\pm R \,e^{\pm \theta}$.  The Rindler metric is
$$
ds^2 = - R^2\,d\theta^2 + d R^2 \, .
\eqn\rinmet
$$
The Rindler Hamiltonian is conjugate to the time variable $\theta$
and is simply the generator of Lorentz boosts in Minkowski space.
At $t=0$,
$$
H_{\rm R} = \int_0^\infty dx\, T^{00}(x) \,x \>.
\eqn\rinham
$$
The Rindler Hamiltonian only acts on a Hilbert space defined in
terms of fields living on the positive $x$-axis and a second Hilbert
space is needed in order to describe degrees of freedom on the
negative $x$-axis.  The Rindler Hamiltonian has a set of eigenvectors
including a ground state.  However, this ground state is not to be
identified with the physical vacuum of Minkowski space, which has
non-vanishing correlations between degrees of freedom on the positive
and negative $x$-axis.  These correlations are strongest at short
distances and the Rindler ground state deviates violently from the
physical vacuum very close to the origin.  In fact, the expectation
value of the Minkowski energy density diverges in the Rindler ground
state at that point.  Since the Minkowski vacuum is defined on the
full $x$-axis it can not be a pure Rindler state but is rather a
thermal density matrix, as is well known [\unruh,\unwald].

An analogous situation arises when a black hole is described in
Kruskal coordinates.  In that case the ``Rindler ground state'' is
singular at the horizon.  It describes a matter field configuration
with divergent energy density which is expected to react strongly
back on the geometry at the horizon and therefore this state is
usually not taken seriously.  It nevertheless remains an open
question whether there exists a corresponding zero temperature
quantum state when gravitational corrections have been taken into
account.  It turns out that such a configuration can be found in
our 1+1 dimensional model, as we shall shortly see, and we expect
one to exist in 3+1 dimensions also.  In fact, this is presumably
the only static macroscopic field configuration in quantum gravity
with a given mass.  A finite temperature black hole in thermal
equilibrium with its surroundings is not consistent because the
thermal bath filling space would have infinite total energy.  Of
course, real black holes with finite Hawking temperature are not
described by a static geometry but are formed in gravitational
collapse.  Hawking's original calculation showed that gravitational
collapse results in a black hole with a non-singular horizon and
outgoing radiation with a thermal spectrum, but the black hole is
not in thermal equilibrium with the rest of space.  Whilst it is
well established that black holes are formed with non-vanishing
Hawking temperature and emit radiation it is very unclear how they
subsequently evolve.  It is an intriguing possibility that the
back-reaction of Hawking radiation on the geometry could cause the
evaporating black hole to eventually settle into one of these massive
zero temperature remnants, and we see no reason to reject this out of
hand.  We will, however, be able to show that this scenario is not
realized in actual gravitational collapse, at least not in our 1+1
dimensional toy model.

Let us first consider the energy momentum of quantum mechanical matter
in a static black hole background ignoring back-reaction on the metric
and dilaton.  The static black hole solution \static\ written in
asymptotically Minkowskian coordinates $\tau,\zeta$ is
$$\eqalign{
e^{-2\rho} =\,& 1+ {M\over \gl} e^{-2\gl \zeta} \,,  \cr
e^{-2\phi} =\,& e^{2\gl \zeta}
\bigl(1+{M\over \gl}e^{-2\gl \zeta}\bigr)\,.\cr}
\eqn\stat
$$
Far away from the black hole the metric is nearly flat and $\phi$
approaches the linear dilaton background.  In Euclidean space this
solution is easily seen to be a coordinate transformation of Witten's
semi-infinite cigar [\witten], the cigar tip being at
$\zeta\rightarrow -\infty$.
The Hawking effect can be described
in terms of the conformal anomaly [\cghs,\chrful], and it is
convenient to express it in terms of the $Z$ field in \zaction .
The curvature acts as a source term in the $Z$ equation of motion,
$$
\nabla^2Z = -\sqrt{N\over 24} R \,.
\eqn\zeom
$$
The cigar is flat far away from the horizon and one finds a static
$Z$ field linear in the spatial coordinate,
$$
Z \simeq \sqrt{N\over 6}  \, \zeta \,.
\eqn\linzet
$$
Inserting this into \zemt\ gives a constant energy density,
${N\over 24}$, of Hawking radiation.
The coefficient $\sqrt{N\over 6}$ only involves the
integrated curvature, or Euler number, of the cigar end, making it
clear that Hawking temperature is independent of mass in 1+1
dimensions.  There also exists a solution for $Z$ which has vanishing
energy density far away but that requires a point source of $Z$
at the cigar tip and this causes the energy density to diverge at the
horizon.  This singular configuration is the zero temperature
``Rindler ground state''.\foot{Once we are willing to entertain such
a singularity the temperature in the asymptotic region can of course
have any value.}

Neither the thermal black hole nor the zero temperature one remain
consistent solutions when  back-reaction is taken into account.  As
we mentioned before a thermal static configuration can not be
consistent for it would have infinite total energy.  On the other
hand it is reasonable to look for a zero temperature static solution
of the CGHS-equations with some finite mass (or ADM-energy).

Static fields $\phi,\rho$
only depend on the spatial variable $\zeta$ and satisfy
the following set of ordinary differential equations in conformal
gauge,
$$\eqalign{
\phi '' =&\, (1{-}{N\over 24}e^{2\phi}) \, \rho '' \, ,   \cr
   (1{-}{N\over 12}e^{2\phi}) \,\phi '' =&\,
  2\,(1{-}{N\over 24}e^{2\phi}) \,
  (\phi '^2 + \lambda^2 e^{2\rho}) \, ,  \cr}
\eqn\steom
$$
along with one constraint which can be written,
$$
\phi '^2 -\rho '\phi ' +{N\over 48} \rho '^2 - \gl^2 e^{2\rho} = 0\,.
\eqn\stcon
$$
The classical static black hole \stat\ is a solution of the equations
without the $Ne^{2\phi}$ terms.  Far away from the black hole the
correction terms tend to zero and a given solution of the full
equations matches onto a classical one with the same ADM-energy.
Unfortunately we have not been able to find the general solution to
the full set of non-linear equations \steom\ in closed form, but they
can easily be handled numerically.  In figures 5 to 6 we display some
results of numerical integration.   We plot both the value of
$e^{2\phi}$ and the scalar curvature as a function of spatial
position for af number of solutions of \steom\ with different
black hole mass.  Figure~4 graphs the coupling
strength and curvature {\it vs.} position in the
classical solution \stat\ for comparison.

The figures clearly demonstrate the behavior alluded to at the
beginning of this section.  In particular, we see that the geometry
and dilaton approach a classical black hole solution at large distances
but before the classical horizon is reached the curvature reaches a
maximum and the decreases to $-\infty$.  Simultaneously the coupling
strength $e^{2\phi}$ reaches a maximum and then tends to zero as the
curvature tends to $-\infty$.  The maximum coupling is always less
than the critical value, ${N\over 12}e^{2\phi_{max}}<1$, but
approaches this value as $M\rightarrow 0$.  The curvature reaches
a maximum positive value before it dips down to $-\infty$.  As
$M\rightarrow 0$ this maximum positive curvature diverges but
at the same time the size of the region which deviates from the
classical solution goes to zero.

The solutions found above are the only static solutions of the
CGHS-equations.  As expected, no solutions exist which describe a
static black hole in thermal equilibrium with its surroundings.

A natural question to ask is whether a black hole formed in a
physical process such as collapse can evolve into a massive zero
temperature object described by these static solutions.  We will
now show that the CGHS-equations do not have solutions in which
this occurs.  That may not prevent quantum mechanical tunneling into
a massive zero temperature configuration at some stage in the
evolution.  However, we expect that to be heavily suppressed at least
until the black hole has lost almost all its mass.

We begin by noting that in the gravitational collapse of a shock
wave the dilaton field increases monotonically on the infall line
$x^+=x^+_0$, until it hits the singularity at
${N\over 12}e^{2\phi}=1$.
We will prove that this continues to
hold on every line of constant $x^+$, {\it i.e.} that it is not
possible for the dilaton field to develop a local maximum at
${N\over 12}e^{2\phi}<1$ as is found in all $M>0$ static solutions.
We know that on each such line $e^{2\phi}$ varies continuously from
zero, as $x^-\rightarrow -\infty$, to ${12\over N}$ at the singularity.
(Recall that the singularity is a space-like curve which asymptotically
approaches the light-like line, $x^-=H$, as shown in figure~2.)
Thus if a local maximum occurs it must be accompanied by a local
minimum closer to the singularity, as illustrated in figure~7.
Call the points where the extrema occur $x^-_1(x^+)$ and
$x^-_2(x^+)$.  At these points $\phim = 0$ and from the second
equation of motion in \eom\ it is seen that $\phipm > 0$ there.
It follows that the two curves $x^-_1(x^+)$ and $x^-_2(x^+)$ must
approach one another as $x^+$ increases.  However, since the $\phi$
is monotonic on the infall line there must be a smallest value of
$x^+$ where the extrema first occur and at which $x^-_1=x^-_2$.
Since the equations of motion prevent $x^-_1$ and $x^-_2$ from
separating it follows that $\phi$ must remain monotonic.

The local maxima of $\phi$ as a function of the spatial variable
$\zeta$ are also local maxima when viewed along constant $x^+$
lines.  Therefore the CGHS-equations do not permit infalling
matter to evolve to a massive remnant.  The only static solution
which has monotonic $\phi$ is the $M=0$ solution.  It seems highly
plausible that the asymptotic endpoint of 1+1 dimensional black
hole evolution is the $M=0$ state.
Late in the evolution the system is expected to become highly
quantum mechanical and the above argument does not rule out the
possibility of forming light stable remnants.

The approach to the vacuum state, as described by the CGHS-equations,
is subtle and non-uniform.  We believe that along any line of
constant $\phi$ in the physical region, ${N\over 12}e^{2\phi}<1$, the
evaporating solution will eventually tend to the vacuum configuration.
However, this cannot occur until the apparent horizon has receded
beyond that value of $\phi$.  As $x^+$ increases there will always
be a small but diminishing region which remains inside the apparent
horizon.  In this region the solution is qualitatively different
from the vacuum, for example constant $\phi$ contours are space-like
there, but any given $\phi$ contour will eventually emerge outside
the horizon and become vacuum-like.  It is unclear how much
physical meaning should be attached to the Liouville region,
${N\over 12}e^{2\phi}>1$.  For our purposes the singularity
can be taken as a boundary of spacetime.

\chapter{The Fate of Quantum Information}

If stable remnants do not store the information that falls into
a black hole then the only way to have a unitary quantum theory is to
find a complete correlation between the state of the infalling matter
which formed the black hole and the final state of the Hawking
radiation.  The fact that the final radiation is approximately thermal
is in no way inconsistent with this.  However, it is important to
exhibit a mechanism for transferring the infalling information to
the outgoing particles.  Consider figure~8 which depicts an infalling
system with some complexity forming a black hole and subsequently
radiating away its energy.  In order that the Hawking radiation carry
away all the information it is necessary that the infalling matter has
had all its information eliminated before it crosses the global event
horizon at $x^-=H$.  If the infalling matter has any memory of the
initial state beyond this point then the Hawking radiation cannot be in
a pure state since it would still be correlated to the infalling matter
which has passed inside the event horizon.
It is not sufficient for the infalling matter to imprint its state on
the outgoing radiation like DNA imprinting itself on RNA.  The
original information in the ``DNA'' must also be destroyed if the
``RNA'' is to be in a pure state.  At first sight this seems
obviously impossible because the $f$-waves satisfy free field
equations of motion and any structure that they have is maintained
all along the infall line.  In fact a free massless left-moving
$f$ field has an infinite number of conserved quantities which
characterize the exact shape of the wave.

On the other hand, since all the energy of the system is emitted
as Hawking radiation the total energy of the infalling matter must
be zero when it arrives at $x^-=H$.  One might wonder whether
all the apparent information carried by the $f$ fields is real.
At the moment we have no definitive answer to this important
question but we can exhibit a similar system in which the apparent
information in an infalling free field is illusory and in which
all real information is radiated in an analog of Hawking radiation.

Consider the 1+1 dimensional Schwinger model\foot{1+1 dimensional
electrodynamics has been considered in
a different context of black hole physics in reference [\alstr].}
with a coupling constant which depends on position like
$$
g^2 = e^{2x} \,,
\eqn\coupl
$$
where $x$ is the spatial coordinate.  This can be thought of as
electrodynamics in the linear dilaton vacuum.  The Lagrangian is
$$
L = - {1\over 4 g^2(x)} F_{\mu\nu}F^{\mu\nu}
    + i \bar\psi \gamma^\mu(\p_\mu-iA_\mu)\psi \,.
\eqn\schlag
$$
The analog for vector fields of the conformal gauge is the
light-cone gauge $A_-=0$.  In this gauge the Lagrangian becomes
$$
L = - {1\over g^2(x)} (\p_-A_+)^2
    + i \bar\psi_L {\p \over \p x^-}\psi_L
    + i \bar\psi_R ({\p \over \p x^-}-iA_+)\psi_R \,.
\eqn\lclag
$$
The field $\psi_L$ describing ``infalling'' fermions appears to
be a free field.  A message could be encoded in sequence of particles
and antiparticles of this type.  Naively, any such information in
$\psi_L$ is conserved and penetrates indefinitely into the strong
coupling region.  However, this is incorrect.  To see this let us
integrate out the $\psi$ field.  As is well known this contributes
a one-loop correction to the effective action of $F_{\mu\nu}$, which
is given by,
$$
{1\over 8\pi} \int d^x \,
\epsilon^{\mu\nu}F_{\mu\nu} \,{1\over \nabla^2}\,
            \epsilon^{\gl\sigma}F_{\gl\sigma}  \>.
\eqn\fanom
$$
Note the similarity with the Liouville action \nonloc\ obtained by
integrating out the matter fields $f_i$.  This action is non-local
but we can introduce a $Z$ field as before.  The effective Lagrangian
then becomes
$$
L = {1\over 2} \p_\mu Z \p^\mu Z +
    {1\over \sqrt{4\pi}} \epsilon^{\mu\nu}F_{\mu\nu}\,Z
    - {1\over 4 g^2(x)} F_{\mu\nu}F^{\mu\nu} \>.
\eqn\zeff
$$
Now integrate the gauge field $A_\mu$ to obtain
$$
L_{\rm Z} ={1\over 2} \p_\mu Z \p^\mu Z+{g^2(x)\over 2\pi}\,Z^2 \>.
\eqn\zlag
$$
This is the bosonized form of the Schwinger model in which the
axial anomaly appears at the classical level.  In terms of $Z$
the charge current is given by $\epsilon^{\mu\nu}\p_\nu Z$.

It is evident from \zlag\ that information cannot penetrate
indefinitely into the strong coupling region $x\rightarrow -\infty$.
The effective mass of the bosonizing field $Z$ grows to infinity
with the coupling so that any incoming $Z$ wave of finite energy
is completely reflected.  This simply indicates the fact that
electric current cannot flow into the infinitely strongly coupled
region.

The possibility of information penetrating the strong coupling
region depends in detail on the model.  For example if more than
one fermion species carries charge then bosonization reveals one
``massive'' boson $Z$ with mass $\sim g^2(x)$ and a set of
massless bosons which can carry information toward $x\rightarrow
-\infty$.  On the other hand if all currents are independently
gauged then all the boson fields are reflected.

It is not clear whether or not a theory can be constructed in which
information is always drained out of the $f$ fields before they slip
behind the global event horizon.
It may be that an arbitrary theory, for example one which has
global symmetries, cannot satisfy this.  Perhaps it is only in some
theories, or even only one, that black hole evolution can be made
consistent with conservation of information.

A possible candidate for such a theory in 1+1 dimensions is obtained
if we completely replace the $f$ fields in the CGHS-model
by the field $Z$.  This way the matter carries no information which
is uncoupled to gravity.  The question is whether in such a theory
a configuration on the global event horizon which carries zero total
energy must also carry no information.  For example if $T^m_{++}$ in
our expression for the energy \mint\ were a positive quantity as it
would be in a classical theory then vanishing energy would require
the $Z$ field be trivial on the event horizon and therefore carry no
information.  Unfortunately energy density does not have to be positive
in quantum theory.

To begin to address this question consider an incoming configuration
described asymptotically at large
negative $x^-$ by some left-moving positive energy
density $T_{++}(x^+)$.   It is always possible to find a $Z$ field
which carries the same energy distribution,
$$
T_{++}(x^+) = {1\over 2} \nabla_+Z\nabla_+Z
              + \sqrt{N\over 24}\,\nabla_+^2Z  \,.
\eqn\zpp
$$
Along a given line of fixed $x^-$ one can find a coordinate
transformation $x^+\rightarrow y^+(x^+)$ which renders the metric
flat along that line.  The new coordinates are still conformal and
\zpp\ can be written as
$$
T_{++}(x^+) = {24\over N} e^{-\sqrt{N\over 24}Z}
       \bigl({\p \over \p y^+}\bigr)^2 e^{\sqrt{N\over 24}Z}
\eqn\zppy
$$
along the line in question.  If we write
$\psi = e^{\sqrt{N\over 24}Z}$ then \zppy\ takes a familiar form.
It is the time independent Schr\"odinger equation for a zero energy
state in a potential $V(x^+)= {N\over 24} T_{++}(x^+)$.  We can impose
a boundary condition that $Z=0$ or $\psi = 1$ in the linear dilaton
vacuum at $x^+<x^+_0$.  If we assume that $T_{++}$ vanishes for large
enough $x^+$ then the solution of the Schr\"odinger equation will be
linear in that region.  If the sign of the linear term is negative
then $\psi$ will have to go through zero at some point, and
$Z$ becomes singular and ill-defined there.  The sign depends on the
potential $V$.  For example, if
$\int V(x^+)\,dx^+ \leq 0$ a zero energy solution of the Schr\"odinger
equation always has a node somewhere.  This means that an incoming
asymptotic configuration with negative total energy would necessarily
be singular and the only non-singular zero energy state has
$T_{++}=0$.

In the case at hand it is not the incoming total energy which vanishes
but rather the energy on the global event horizon.  This condition
does not appear to be strong enough to force $Z$ to be vacuum-like
along the entire global horizon.  To see that introduce a $y^+$
coordinate system with flat metric on the global event horizon,
$x^-=H$.  The expression for the (vanishing) total energy \mint\
becomes
$$
\int_0^\infty dx^+\, \phim \sqrt{1-{N\over 12}e^{2\phi}}
           [\psi^{-1} ({\p \over \p x^+})^2 \psi]  = 0 \,.
\eqn\zecond
$$
In general the weight factor $\phim \sqrt{1-{N\over 12}e^{2\phi}}$ is not
constant and that prevents us from concluding that $Z$ must be trivial.
In fact the weight factor increases with increasing $x^+$ and this
might allow $T_{++}$ to be positive on and near the infall line but to
take a smaller compensating negative value at larger $x^+$.
Unless the back-reaction on the geometry provides more stringent
constraints on $T_{++}$ on the global horizon then even in this
``minimal'' theory all information will not be drained out of the
incoming $Z$ field.

\noindent
\undertext{Acknowledgement:}
We thank J.~Russo for numerous discussion and collaboration.  We have
also benefitted from discussions with M.~Alford, B.~Birnir, C.~Callan,
S.~Giddings, G.~Horowitz, J.~Preskill, M.~Srednicki, and A.~Strominger

\refout
\end